\documentclass[twocolumn,twoside,slac_two]{revtex4}
\usepackage{graphicx}
\usepackage{fancyhdr}
\usepackage{subfigure}
\usepackage{mdwlist}
\begin{document}

\title{Charged cosmic rays: \\ a review of balloon and space borne measurements}

\author{Pier Simone~Marrocchesi}
\affiliation{Dept.$\,$of Physical Sciences, Earth and Environment, Univ.of Siena, v.Roma 56, 53100 Siena (Italy) \\
                and INFN Sezione di Pisa, Largo B. Pontecorvo 3, 56127 Pisa}  

\begin{abstract}
Current experimental data on cosmic-ray charged leptons are briefly reviewed including measurements of the positron fraction, electron and positron spectra and inclusive electron+positron data. Precision measurements by PAMELA and AMS-02 of the positron spectrum and its striking difference from the electron spectrum have prompted several theoretical speculations. In order to discriminate among different interpretations of the “positron anomaly”, a major step is needed to achieve an accurate direct measurement of the shape of the inclusive electron spectrum beyond 1 TeV. Ongoing efforts along this direction by instruments already in orbit and by the recently launched CALET and DAMPE missions are described. 
	A challenging experimental scenario, at variance with the standard paradigm of a single power law spectrum, emerges for the cosmic-ray charged hadrons after the discovery of a discrepant hardening in the rigidity spectra of protons and He in the 200 to 300 GV region  (CREAM, PAMELA, AMS-02) and the observation by AMS-02 of a possible break in the Li spectrum. An additional puzzle comes from the apparent violation of the universality of spectral indices whereby He and proton spectra are progressively hardening above $\sim$100 GV with a similar rigidity dependence, but the He spectrum is harder than proton's.  
	Secondary-to-primary ratios (most notably B/C) and isotope flux ratios provide insight into the subtleties of propagation mechanism(s) and test the internal consistency of the models of acceleration and propagation of cosmic rays in the galaxy.
	Important complementary information on the abundance of nuclei heavier than iron comes from dedicated balloon-borne instruments as SuperTIGER or space-based missions as ACE at L1 that provided the first measurement of a primary `cosmic-ray clock' with the CRIS instrument. 
	Measurements of the high energy anti-proton spectrum and its ratio to the proton spectrum by PAMELA and AMS-02 and consistency with BESS data at lower energy are briefly discussed. 
	A brief overview of future missions follows, including those ready for launch as ISS-CREAM and others, at different stages of design, as HERD (proposed for the future Chinese Space Station) or balloon instruments including GAPS (detection of low energy deuterons), HELIX (measurement of $^{10}$Be/$^{9}$Be ratio) and HNX (heavy nuclei).
\end{abstract}

\maketitle

\thispagestyle{fancy}

\section{Introduction}

The present scenario of cosmic-ray (CR) phenomenology is in rapid evolution. On the one hand, recent measurements made available mainly by long duration missions in space (PAMELA, FERMI, AMS-02), have provided -- with an unprecedented precision -- data on the functional dependence of some important parameters (e.g.: spectral indices versus rigidity) with a significant improvement on previous measurements that suffered from statistics limitations (and systematic uncertainties) due to their relatively short exposure.
On the other hand, the new data have shown some unexpected results and prompted a number of open questions. \\
This paper refers to an invited review talk given in September 2016 at the ECRS in Turin. Since then, new important results by the AMS-02 collaboration were published (including the new measurement of the B/C ratio \cite{AMSbtoc}) or presented in a public seminar \cite{Ting2017}. Therefore, part of the information contained in the present report has to be integrated with new material that can be found in the references provided herein. \\
Due to the obvious limitations, both in the time available for the talk and the space allowed for the written summary, a number of interesting topics had to be omitted or just briefly mentioned without further details. The author apologizes for the inevitable incompleteness of this report. It should be regarded as a very concise summary of the major highlights in CR phenomenology rather than a comprehensive overview of the state of the art in the field.  \\
The paper is organized as follows.
Sections \ref{s:electron} and \ref{s:positron} summarize the latest results of the measurements of electrons and positrons, while the most recent measurements of proton and helium are discussed in section \ref{s:pHe}, followed by 
results on heavier nuclei in section \ref{s:nuclei}. 
Anti-proton measurements are only briefly mentioned in section \ref{s:antiproton}.  \\
A review of recently launched or upcoming cosmic-ray missions, as well as a few examples of experiments at proposal stage are presented in section \ref{s:futurexp}.  

\section{The Electron Spectrum}
\label{s:electron}

In the last decade, an interesting scenario emerged from the new electron measurements provided by ballons, space missions and ground-based instruments.
The balloon experiment ATIC-2 reported the observation of an excess in the inclusive $e^++e^-$ spectrum, in the energy region between 300 and 800 GeV \cite{ATIC-ele}, a result that was not confirmed by FERMI and HESS.
The FERMI spectrum in \cite{FERMI} follows a single power-law (SPL) with spectral index -3.08$\pm$0.05 and it is consistent with the SPL spectrum obtained by the ground-based H.E.S.S. Atmospheric Cherenkov Telescope with index 3.0 $\pm$ 0.1(stat.) $\pm$ 0.3(syst.). The latter features a rapid steepening above 1 TeV \cite{HESS}. \\

\begin{figure} [!h]
\begin{center}
\subfigure[]
{
\includegraphics[height=6.6cm, width=5.4cm,angle=270]{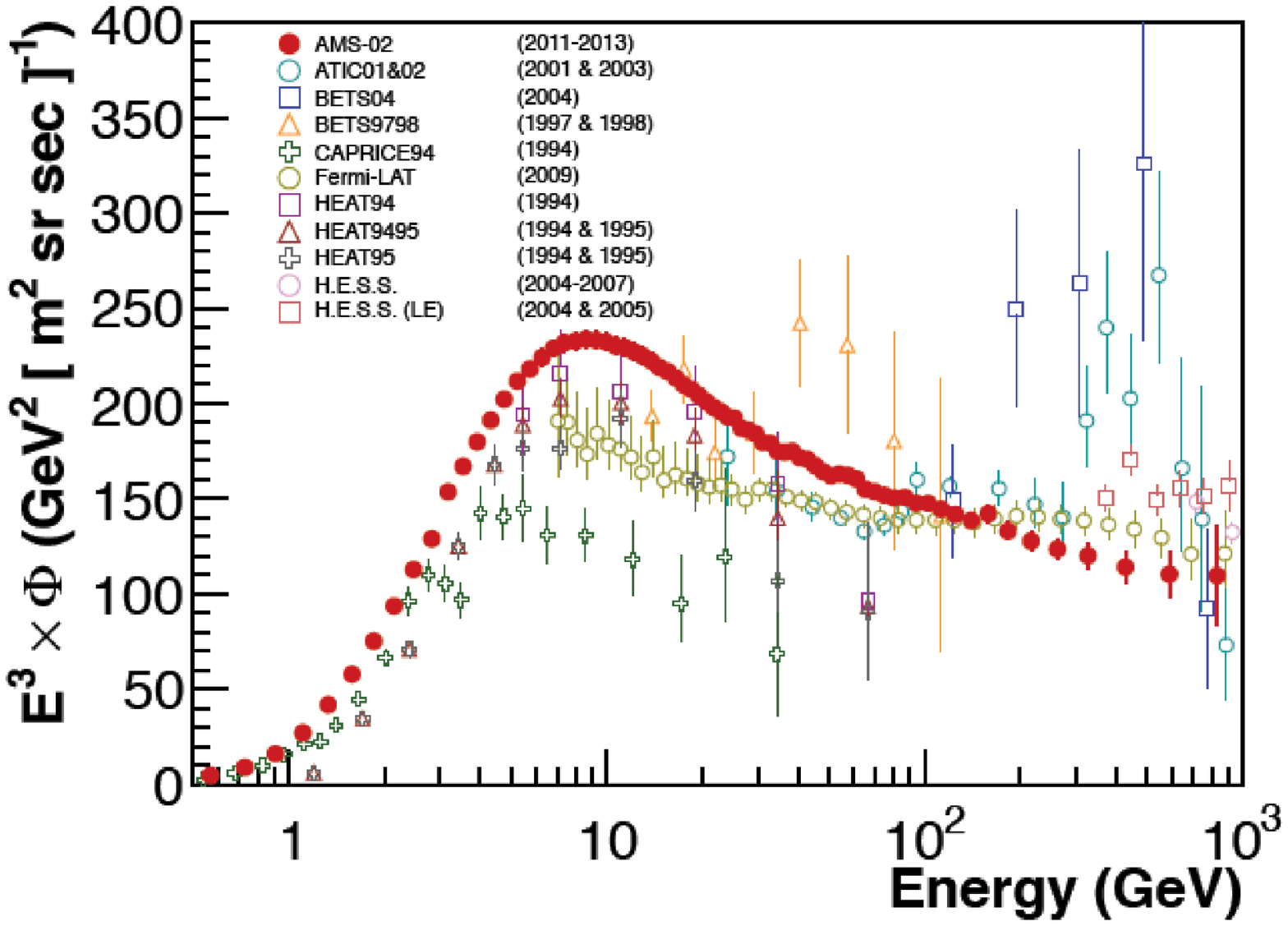}
\label{eleposComp}                            
}
\hspace{10mm}
\subfigure[]
{
\includegraphics[height=6.6cm, width=5.4cm,angle=270]{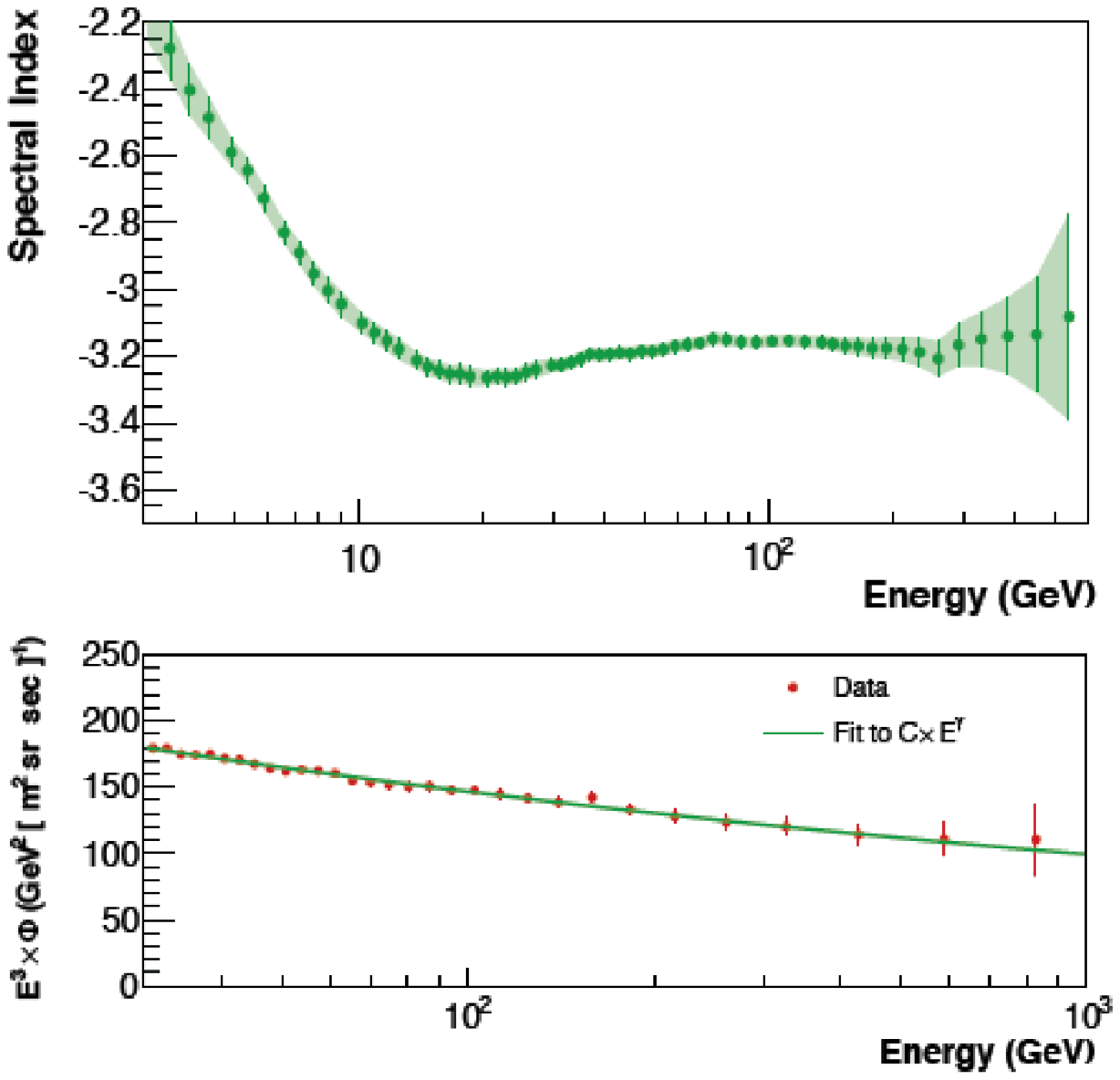}
\label{eleposfit}                            
}
\caption{(a) The AMS-02 inclusive $e^++e^-$ flux 
multiplied by $E^3$ as a function of energy $E$, compared with earlier results.  
(b) Top: spectral index as a
function of energy. Bottom panel:
single power-law fit to the AMS-02 $e^++e^-$ flux multiplied by $E^3$ above 30.2 GeV \cite{AMS-allelectron}.
}
\label{elepos}
\vspace{-0.8cm}
\end{center}
\end{figure}

In 30 months of operations, AMS-02 identified $\sim$10.6$\times$10$^6$ $e^++e^-$ events between 0.5 GeV and 1 TeV.
The resulting flux (Fig.~\ref{eleposComp}) is consistent with a single power-law above 30 GeV with spectral index 
$\gamma$ =-3.170$\pm$0.008 (stat+syst)$\pm$0.008 (energy scale), as shown in Fig.~\ref{eleposfit}. No structures in the $e^++e^-$ spectrum \cite{AMS-allelectron} were reported.\\

The energy was measured by the e.m. calorimeter (ECAL) with a good resolution ($\sim$2\% at 300 GeV), while the magnetic spectrometer provided sign-of-charge discrimination and an independent momentum measurement. The latter is expected to match the calorimetric energy for electrons but not for hadrons.  When combined with the ECAL separation power based on shower shape analysis, it provides an overall proton rejection power of about 10$^4$ above a few GeV (for a  90\% $e^\pm$ ECAL selection efficiency). The independent proton rejection power provided by the TRD detector at 90\% $e^\pm$ efficiency is $\sim$2$\times$10$^4$ at 30 GeV, rapidly decreasing above.\\

\begin{figure}
\begin{center}
\vspace{5mm}
\subfigure[]
{
\includegraphics[height=6.5cm, width=5.4cm,angle=270]{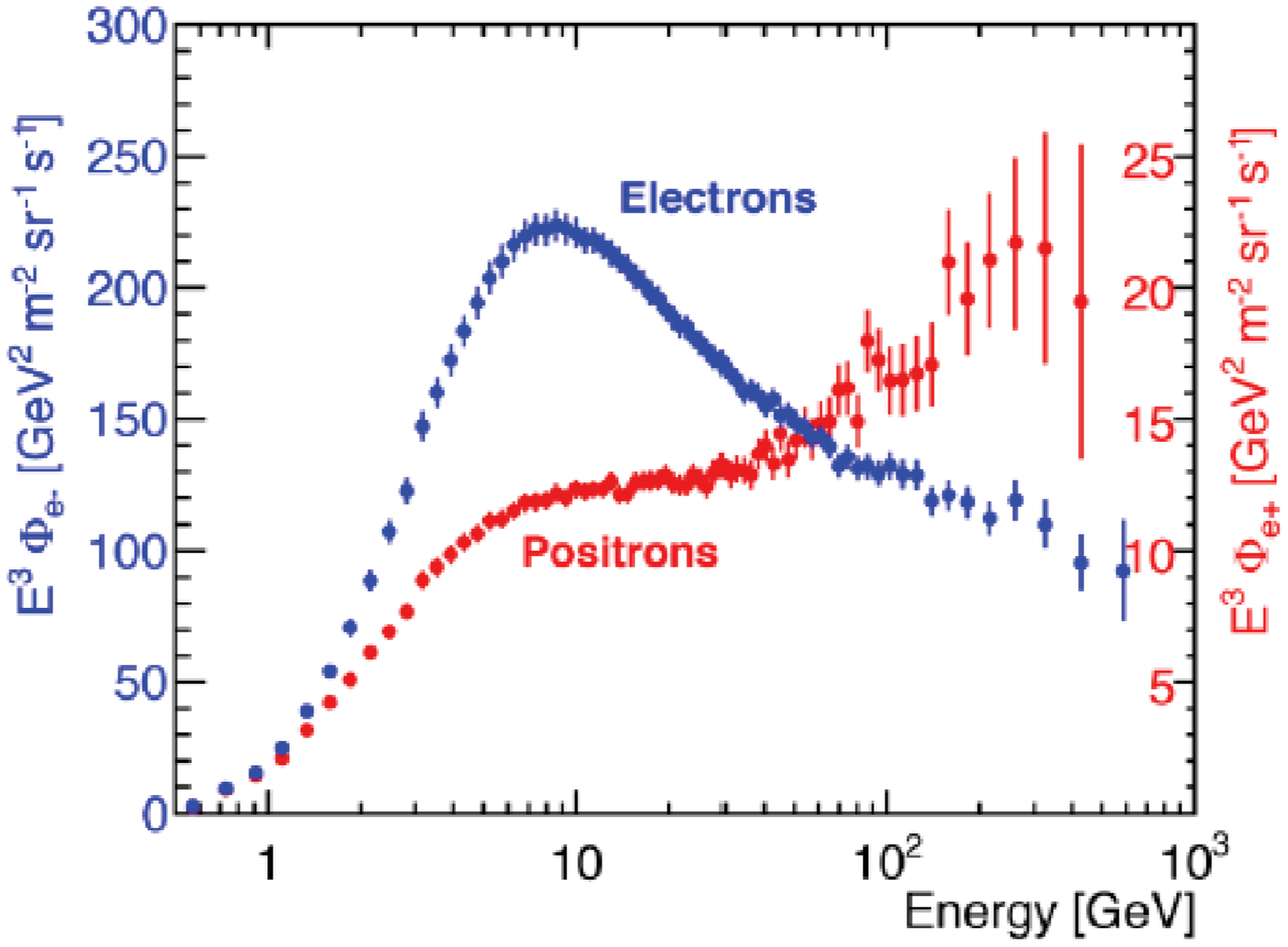}
\label{elepos_comp}                            
}
\hspace{10mm}
\subfigure[]
{
\includegraphics[height=6.2cm, width=5.0cm,angle=270]{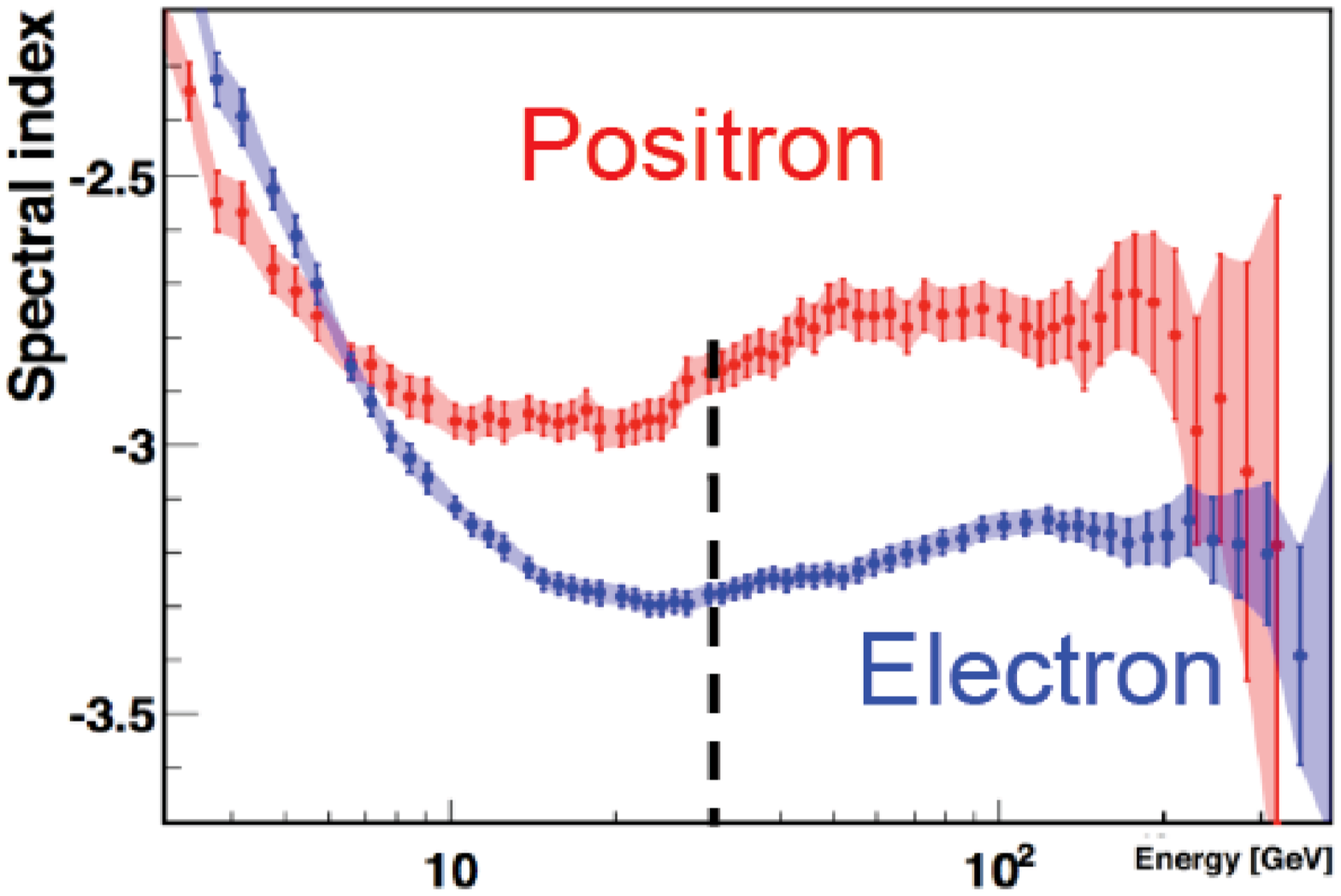}
\label{elepos_spectralindex}                            
}\vspace{-0.3cm}
\caption{(a) $e^-$ and $e^+$ fluxes multiplied by $E^3$ as measured by AMS, 
(b) The spectral indices $\gamma_{e^\pm}$ of the AMS $e^\pm$ fluxes  
 as a function of energy \cite{AMS-electronICRC,AMS-elepos}.
}
\label{figepm}  
\vspace{-0.4cm}
\end{center}
\end{figure}

The exclusive $e^-$ spectrum measured by PAMELA between 1 and 625 GeV shows no significant spectral features in this region within the error and it is well described by a single power-law with spectral index -3.18$\pm$0.05 above 30 GeV \cite{PAMELA-ele}. \\
A precise measurement of the $e^-$ flux was also contributed by AMS-02 (Fig.~\ref{figepm}(a)),
based on the identification of 9.23$\times$10$^6$ $e^-$ events in the energy range 0.5-700 GeV \cite{AMS-electronICRC}.
The collaboration concluded that the spectrum is not consistent with a single power-law above 10 GeV (where the effects of solar modulation are weak) and that the spectral index $\gamma_{e^-}$ shows an energy dependence (Fig.~\ref{figepm}(b)) with a hardening above $\sim$30 GeV \cite{AMS-elepos}.\\

\vspace{-8pt}
\underline{$e^++e^-$ spectrum: the sub-TeV region} \\
The most energetic galactic cosmic-ray (GCR) electrons that can be observed from Earth are likely to originate from sources younger than $\sim$10$^5$ years and located at a distance less than 1 kpc from the Solar System. This is related to the light mass of the electron that allows for large radiative energy losses (by synchrotron radiation and inverse Compton scattering) limiting the propagation lifetime of high energy electrons and consequently the distance they can diffuse away from their source(s).
The electron energy spectrum from 10 GeV to 1 TeV could be the result of the contribution of one or more unresolved sources. Their nature, either as astrophysical objects (e.g. a nearby pulsar) or the product of the annihilation/decay of dark matter particles 
is unclear and is being investigated in the effort to interpret the recent electron measurements that suggest a hardening of the inclusive spectrum in the range 200 GeV - 1 TeV.  The presence of an additional spectral component is also proposed to explain the now established rise of the positron fraction above $\sim$10 GeV, as measured by PAMELA \cite{PAMELA-ele} and extended to the hundreds GeV region by AMS-02 \cite{AMS-elepos}.\\
The observational scenario is complemented by the observations of X-ray and TeV gamma-ray emission from SNRs that provide a clear evidence for the acceleration of electrons in SNR shocks up to energies of about 100 TeV \cite{SNRaccele}.\\

\underline{The TeV region}.
An exciting possibility is that the new measurements of the electron spectrum in the TeV region might result, in the relatively near future, in the direct detection of nearby astrophysical sources of high energy electrons. Since the number of known candidates satisfying the above constraints is very limited, the energy spectrum of electrons might show distinctive structures leading to the identification of the source(s) \cite{Nishimura1980}, \cite{Kobayashi2004}.
In this scenario, the arrival directions are expected to show an anisotropy that increases with energy and hopefully will reach the threshold of detectability.
There are at least 9 candidate Supernova Remnants (SNR) with ages $<$ 10$^5$ years at distances less than 1 kpc from the solar system. Possible contributions to the observed GCR electron spectrum from both distant and nearby sources were calculated. Known candidates that may give a contribution in the TeV region include Vela, Cygnus loop and Monogem in order of strength. Among these, the relatively young Vela ($\sim$ 10$^4$ years) at a distance of ($\sim$ 0.25 kpc) is a very promising
candidate. \\
\noindent The TeV region might as well conceal a completely different scenario
where "nearby" acceleration sources would not be detected and the spectrum would instead roll off at a characteristic cutoff energy.  In this case, the measurement of the "end point" of the electron spectrum can be used to constrain the cosmic-ray diffusion coefficient.\\

At the time of writing, two new space missions CALET and DAMPE have become operational. Both have the capability of providing crucial measurements of electrons above 1 TeV with excellent energy resolution. A brief review of their main characteristics and science goals is provided in section \ref{s:futurexp}.

The TeV region, with its high potential to unveil the presence of local cosmic accelerators, is still largely unexplored.
Thanks to their large collection area, Imaging Atmospheric Cherenkov Telescopes (HESS, MAGIC) proved to be able to 
extend the  $e^++e^-$ spectrum beyond 1 TeV with large statistics, 
at the expense of large systematic uncertainties in the identification of the CR particle that generate the air shower.
In \cite{VERITAS},
 the ground-based experiment VERITAS reported a measurement of the electron spectrum between 300 GeV and 4 TeV.
VERITAS data confirm a cutoff in the spectrum around 1 TeV, first observed by HESS \cite{HESS}, 
above which the spectral index rapidly changes from $-3.2 $ to $-4.1$ approximately.\\

\vspace{-10pt}
\section{The positron spectrum}
\label{s:positron}
The first measurement of the positron fraction $e^+/(e^++e^-)$ to 100 GeV, with significantly better precision than earlier data, was reported by PAMELA \cite{PAMELA-posfrac}. Previous experiments, as the balloon-borne magnetic spectrometers HEAT\cite{HEAT} and CAPRICE \cite{CAPRICE}, as well as the short AMS-01 flight in space \cite{AMS01} had measured the positron fraction up to 30 GeV. They reported a decreasing trend with increasing energy with a small excess around 7 GeV. However, they had limited statistics above 10 GeV and were affected by a large systematic error on the proton background.
On the contrary, PAMELA measurements showed a clear evidence of a rise of the positron fraction above 10 GeV.
This result pointed out for the first time that a purely secondary production, predicting a decrease with energy, was not adequate to describe the data.
The FERMI collaboration confirmed the rise between 20 and 200 GeV.  Not equipped with a magnetic spectrometer and therefore unable to distinguish the sign-of-charge, FERMI exploited the Earth's shadow to separate $e^-$ from $e^+$, though with large systematic uncertainties \cite{FERMI2}.\\


The positron fraction was later measured with high statistics by AMS-02 in the range from 0.5 to 500 GeV\cite{AMS-posfrac}.  It was shown to increase from 10 to $\sim$250 GeV reaching a maximum at 275$\pm$32 GeV. 
A recent update to these measurements \cite{Ting2017} suggest that at higher energies, the positron fraction is actually  decreasing. However,  the data are still statistically limited to provide a reliable discrimination among different theoretical interpretations. At the moment AMS-02 is the only space mission that can provide sign-of-charge discrimination and improve the existing data by extending the current integrated exposure in a long term program of observations on the ISS. \\
The exclusive AMS $e^-$ and $e^+$ fluxes (Fig.~\ref{figepm}(a)) look quite different whereby the positron spectrum is definitely harder than the $e^-$
spectrum. Both spectral indeces show a similar energy dependence, both hardening above $\sim$30 GeV with $\gamma_{e^-} < \gamma_{e^+}$ between 20 GeV and 200 GeV (Fig.~\ref{figepm}(b)). An important feature is that the $e^+$ spectrum is intriguingly similar to the proton's with a spectral index $\gamma_{e^+}$.  As the electron flux is softer than the positron's, the increase with energy observed in the
positron fraction is likely to be caused by the hardening of the positron spectrum and not by a loss of electrons above 10 GeV \cite{AMS-elepos}.\\
The observation of a sharp rise in the positron fraction prompted a number of theoretical models striving to explain the excess of $e^+$, either by including additional sources of high-energy $e^\pm$ or by critically reassessing the secondary $e^\pm$ production from CR interactions with the ISM. 
For the latter case, it was shown, for instance, that by tuning the CR propagation parameters \cite{Blum}, 
or by using inhomogeneous diffusion models \cite{Tomassetti1}, it is possible to reproduce the data under the assumption of a purely secondary production (see also \cite{Cowsik0}). In \cite{Blasi2}, the positron excess is explained taking into account  $e^\pm$  produced as secondaries in the SNRs where CRs are accelerated.\\
As an alternative, astrophysical objects like nearby pulsars, or dark matter (DM) particles have been proposed as possible primary sources of positrons. 
In \cite{ICRCBoudad}, both hypotheses have been tested against the recent AMS-02 data. 
They claim that leptophilic DM is disfavoured, while DM annihilation in quarks and gauge or Higgs bosons would be able to reproduce the positron excess only at the expense of assuming relatively large cross-sections that are not compatible with the constraints provided by other observations.
Instead, a possible explanation based on a single pulsar seems more plausible and five known nearby objects of this kind have been identified that can fit the data.


However, the uncertainties in CR propagation parameters significantly affect the predicted positron rise for both scenarios \cite{ICRCBoudad2}, therefore
strongly limiting at present the possibility of revealing the origin of the excess.
Above 500 GeV the positron fraction energy dependence is expected to be different in two cases of an astrophysical or a DM origin of the $e^+$ excess.
Furthermore, in the presence of a small number of pulsar sources, a dipole anisotropy of order 1\%  should be observed at hundreds of GeV.  
So far, AMS-02 found an $e^+/e^-$ ratio consistent with the measured isotropy \cite{AMS-posfrac}, while separate anisotropy measurements for electrons and positrons are underway \cite{AMS-anisotropy1}.



\section{Proton and helium}
\label{s:pHe}
Early measurements of proton and helium spectra by the balloon-borne experiments JACEE and RUNJOB 
produced controversial results \cite{RUNJOB,JACEE}. Measurements by JACEE suggested a harder spectrum for He than protons at multi-TeV energies while this was not the case for RUNJOB. They were followed by a new generation of ballon instruments based on electronic detection systems rather than emulsion techniques. CREAM collected 161 days of exposure during 6 flights in Antarctica with an energy reach above 100 TeV and provided evidence that proton and He spectra in the multi-TeV energy region are harder than at lower energies 
($<$100 GeV/n) with a proton spectral index $\gamma_p$= -2.66$\pm$0.02 implying a spectrum steeper than helium's ($\gamma_{He}$= -2.58 $\pm$0.02) in the particle energy range 2.5-250 TeV \cite{CREAMpHe}.  A similar hardening in the spectra of the most abundant primary
heavy nuclei (C, O, Ne, Mg, Si, Fe) was reported by CREAM above ∼200 GeV/n \cite{CREAM-harden}.\\
The orbital magnetic spectrometer PAMELA, first observed that the spectral shapes of $p$ and He are inconsistent
with a single power-law. PAMELA $p$ and He data clearly showed an unexpected spectral hardening occurring between 230 and 240 GV,
with a spectral index variation $\Delta\gamma$ $\sim$0.2 for $p$ and 0.3 for He, respectively \cite{PAMELApHe}.\\
In 2015, the AMS collaboration published their results on $p$ and He spectra. Precision measurements of the two spectra clearly showed a change in the spectral indices with both spectra developing a curvature starting from the same energy region observed by PAMELA with a smooth transition to harder spectra.  In the first four years of operation, AMS-02 collected the impressive figure of 68$\times$10$^9$  CR events. A selection of 300$\times$10$^6$ protons and 50$\times$10$^6$ He nuclei was used to measure the differential fluxes as a function of rigidity in the range  
between 1 GV and 1.8 TV for $p$ \cite{AMS-proton, AMS-protonPRL}, and from 1.9 GV to 3 TV for He \cite{AMS-helium}.

\begin{figure}
\begin{center}
\vspace{2mm}
\subfigure[]
{
\includegraphics[height=6.5cm, width=4.4cm,angle=270]{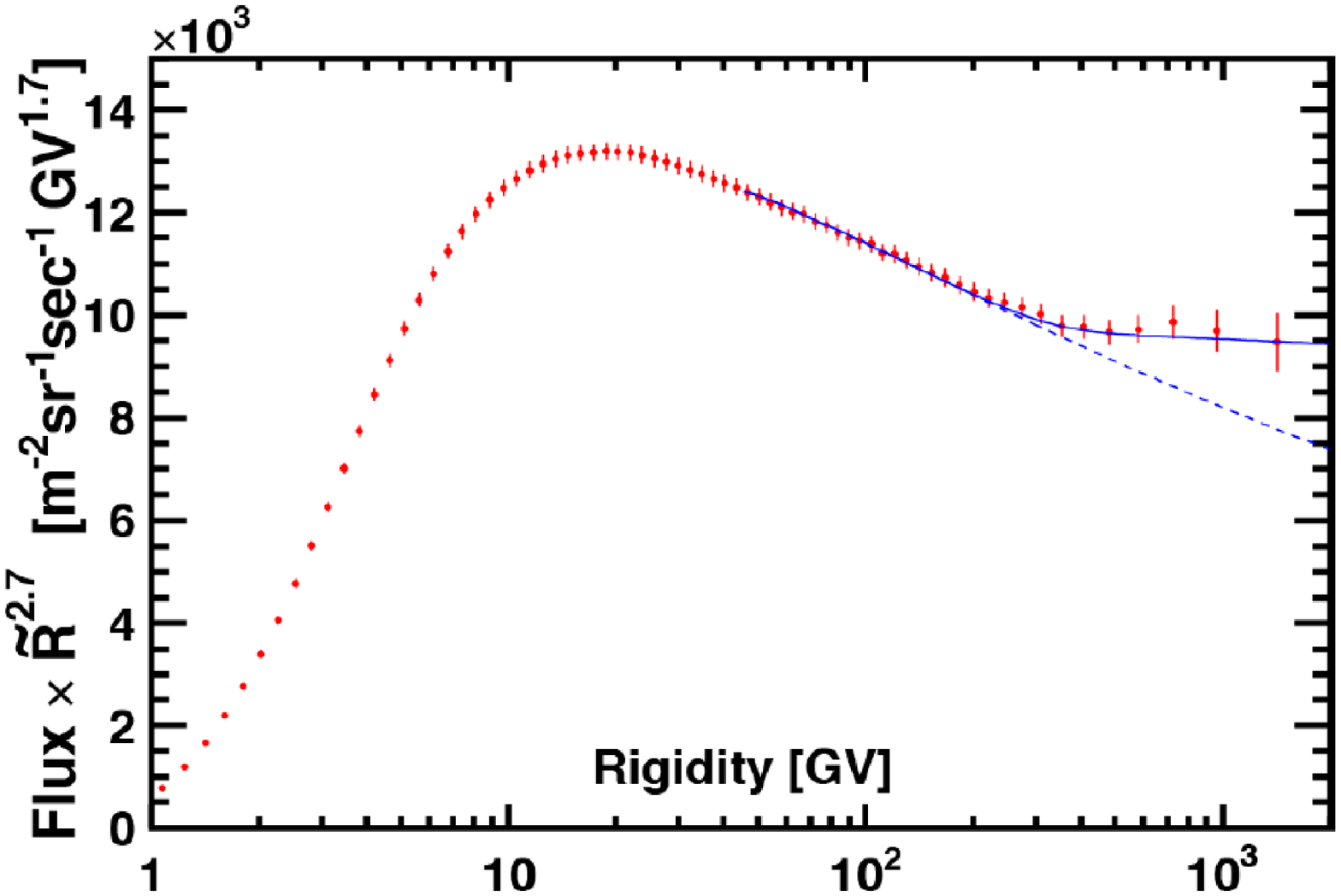}
\label{pfluxa}                            
}
\subfigure[]
{
\includegraphics[height=6.5cm, width=4.4cm,angle=270]{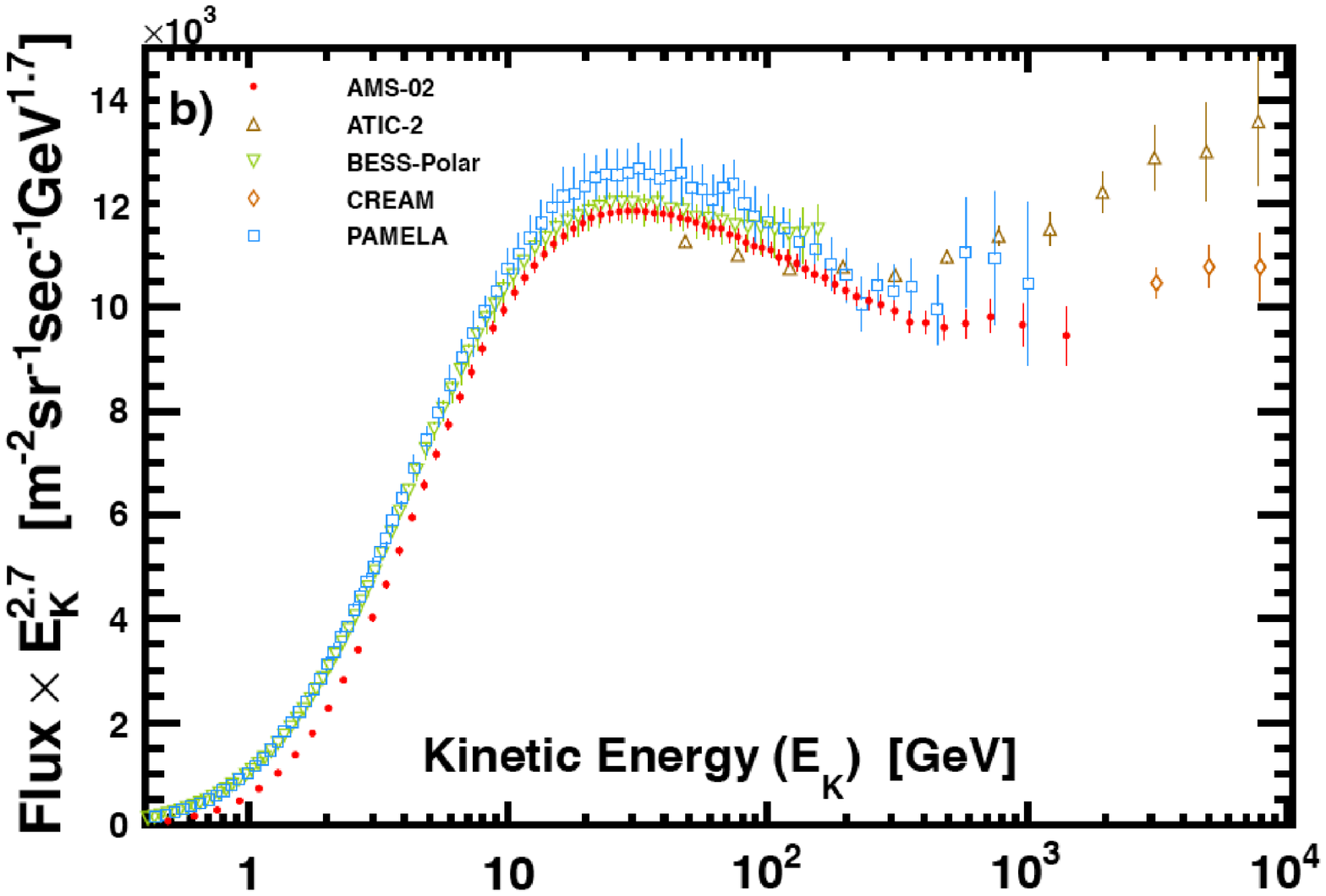}
\label{pfluxb}                            
}\\
\vspace{2mm}
\subfigure[]
{
\includegraphics[height=6.5cm, width=4.4cm,angle=270]{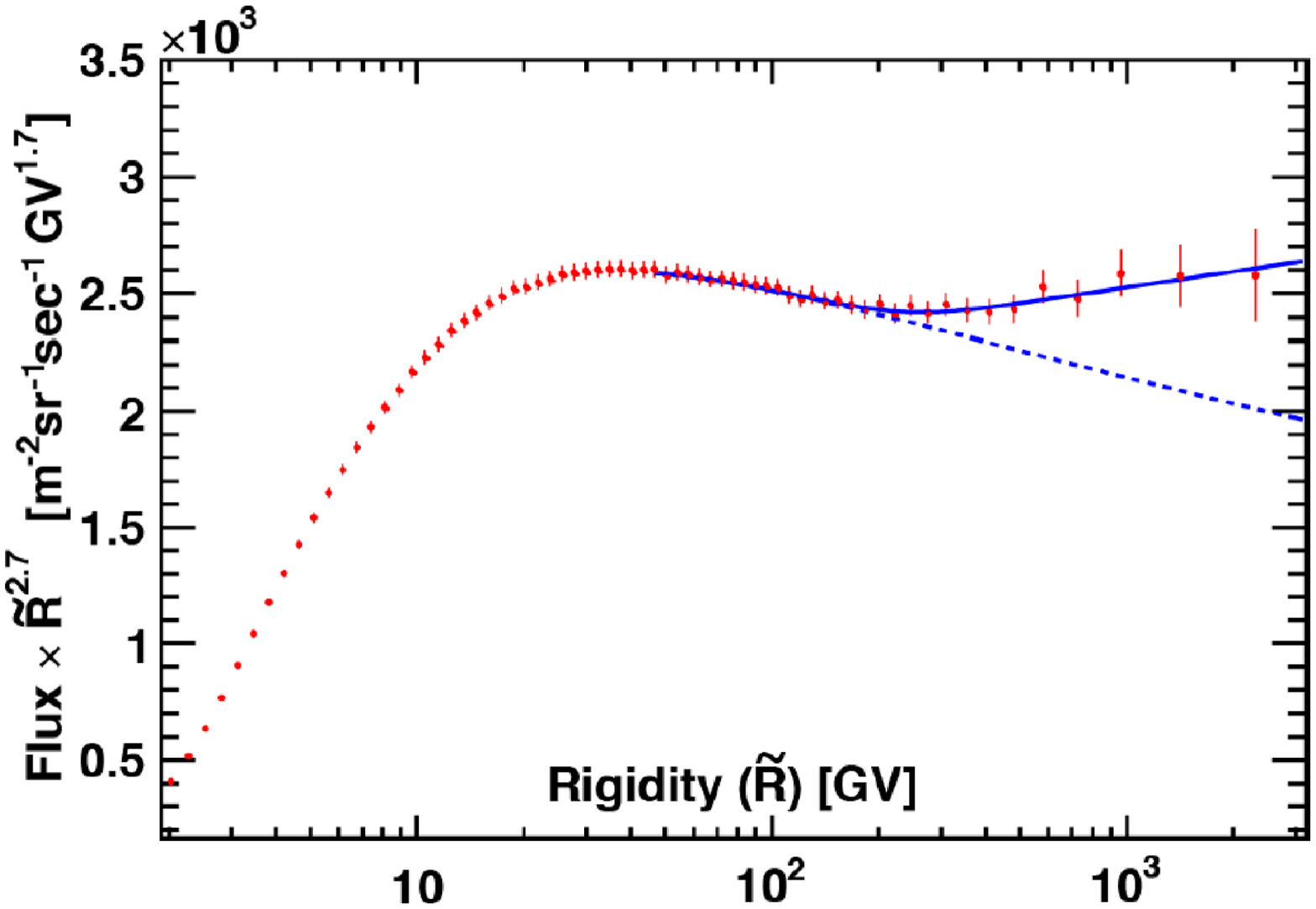}
\label{Hefluxa}                            
}
\subfigure[]
{
\includegraphics[height=6.5cm, width=4.4cm,angle=270]{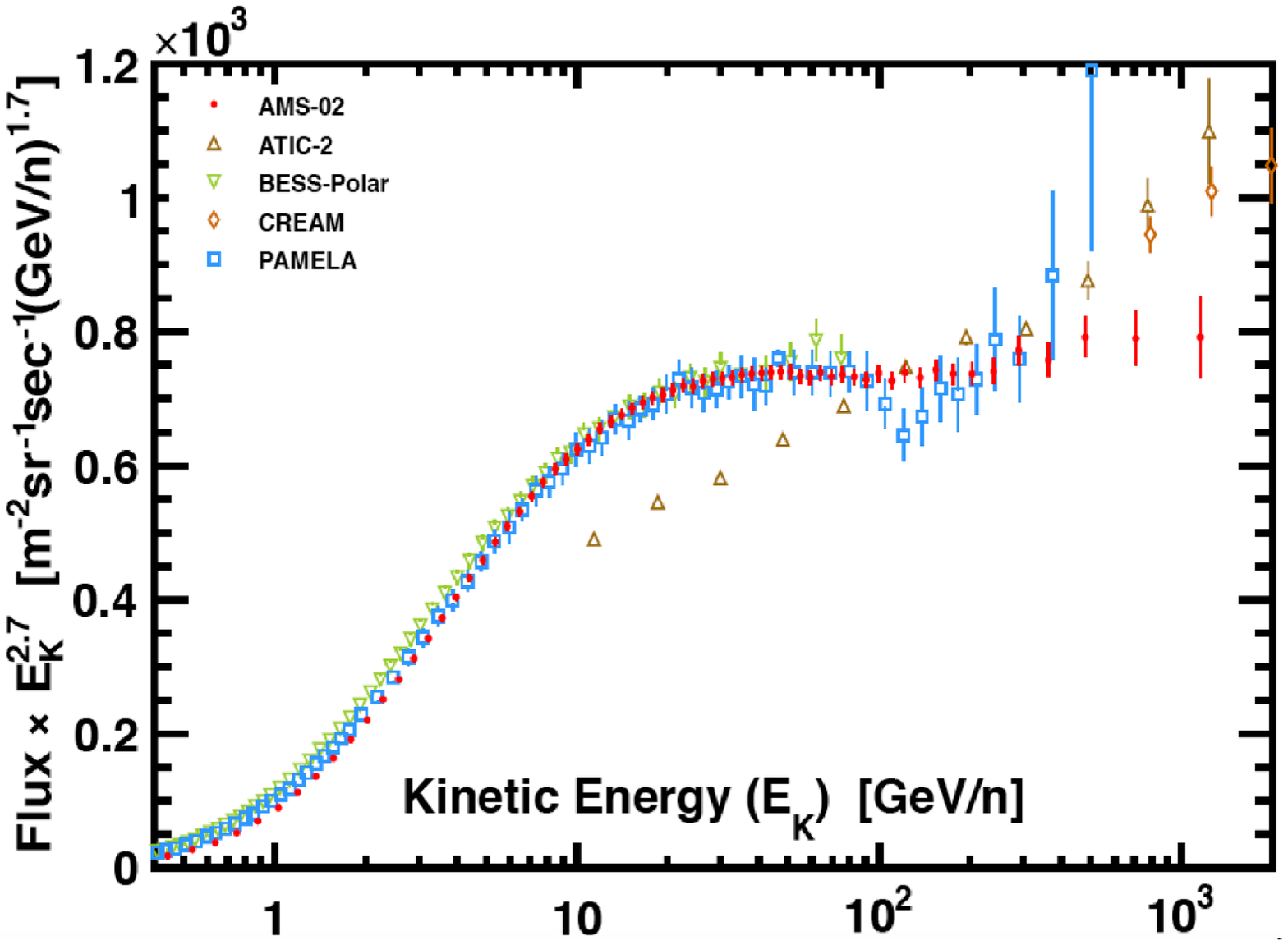}
\label{Hefluxb}                            
}
\caption{AMS (a) $p$ and (c) He  fluxes multiplied by $R^{2.7}$ vs rigidity $R$. The fit of
Eq.~\ref{eq1}  (solid line) and a single power-law  (dotted line) are also shown.
AMS (b) $p$ and (d) He fluxes multiplied by $E_k^{2.7}$  vs kinetic energy per nucleon  $E_k$, compared
with the results of BESS, PAMELA, ATIC and CREAM)\cite{AMS-proton, AMS-helium}.
}
\label{pflux}
\vspace{-4mm}
\end{center}
\end{figure}


The AMS $p$ and He data (Figs.~\ref{pfluxa} and \ref{Hefluxa} ) can be fitted above 45 GV to a double power-law function
\begin{equation}
\Phi = C \left(\frac{R}{\text{45\, GV}}\right)^{\gamma} \left[ 1 + \left( \frac{R}{R_0}\right)^{\frac{\Delta\gamma}{s}} \right]^s
\label{eq1}
\end{equation}
with a smooth transition from a spectral index $\gamma$ below the transition rigidity $R_0$
to $\gamma+\Delta \gamma$ above. 
The fitted spectral hardening for $p$ occurring at $R_0$= 336$^{+68}_{-44}$(fit)$^{+66}_{-28}$(sys)$\pm$1(sol) GV is 
$\Delta\gamma$= 0.133$^{+0.032}_{-0.021}$(fit)$^{+0.046}_{-0.030}$(sys)$\pm$0.005(sol), where the last systematic uncertainty is attributed to 
to the variation of the solar modulation.

For He, $\Delta\gamma$= 0.119$^{+0.013}_{-0.010}$(fit)$^{+0.033}_{-0.028}$(sys)
and $R_0$= 245$^{+35}_{-31}$(fit)$^{+33}_{-30}$(sys) GV.

Proton and He spectra have very similar rigidity dependence, 
but the He spectral index is different with $\gamma_{He}=$-2.780$\pm$0.005(fit)$\pm$0.001(sys) vs. $\gamma_{p} =$-2.849$\pm$0.002 (fit)$\pm^{+0.004}_{-0.003}$(sys) $^{+0.004}_{-0.003}$(sol), respectively.
Furthermore, the detailed study of the variation of $\gamma$ with rigidity, shows that 
the spectral index  progressively hardens above 100 GV for both  species, while the $p$ spectrum remains softer than the He.  

The $p$/He flux ratio can be fitted to a single-power law in rigidity 
above 45 GV with a constant spectral index -0.077$\pm$0.002 (stat)$\pm$0.007 (sys) \cite{AMS-helium}.\\

\begin{figure}
\begin{center}
\vspace{5mm}
\subfigure[]
{
\includegraphics[height=6.0cm, width=8.5cm]{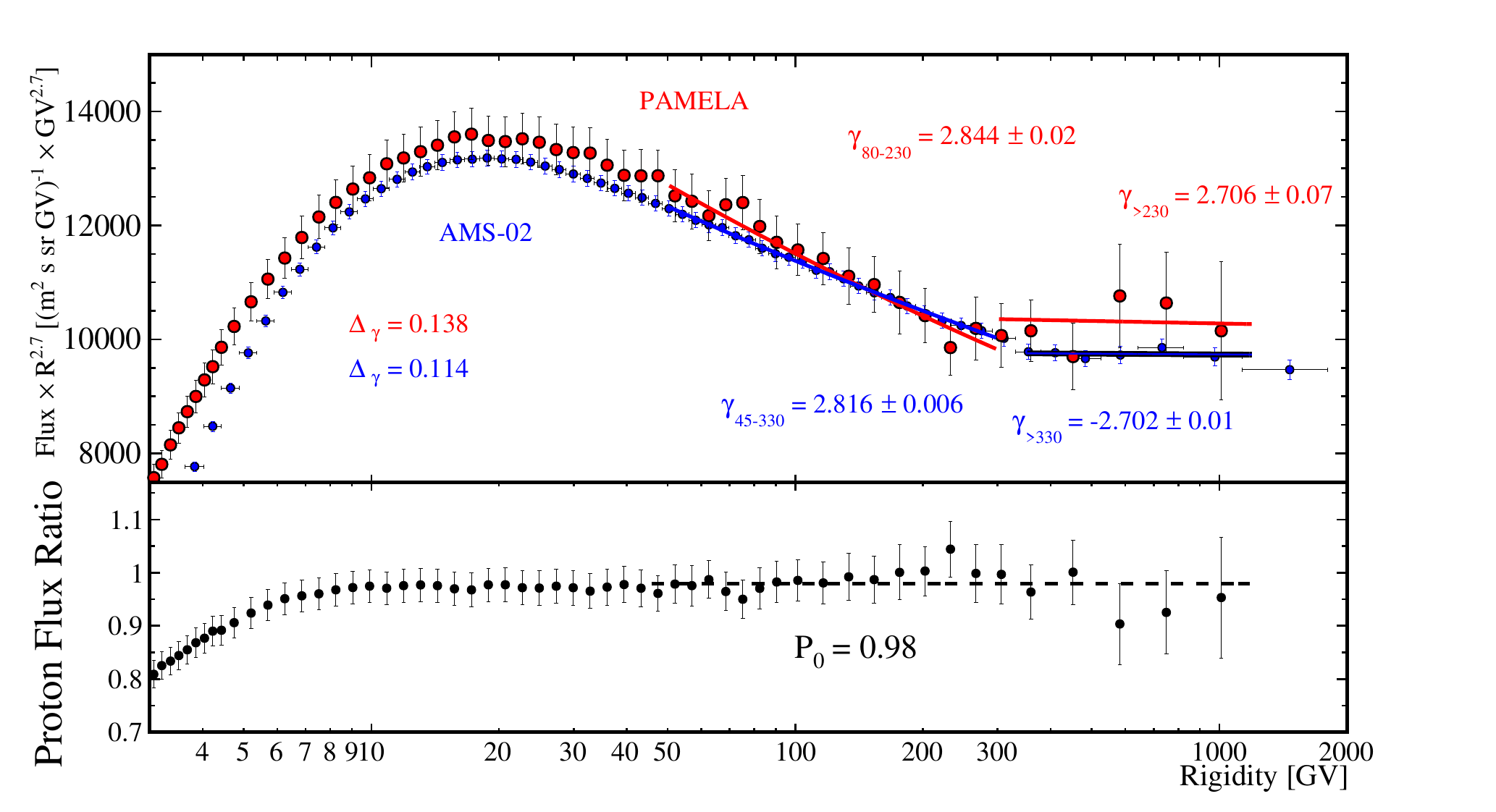}
\label{elepos_comp}                            
}
\hspace{10mm}
\subfigure[]
{
\includegraphics[height=6.0cm, width=8.5cm]{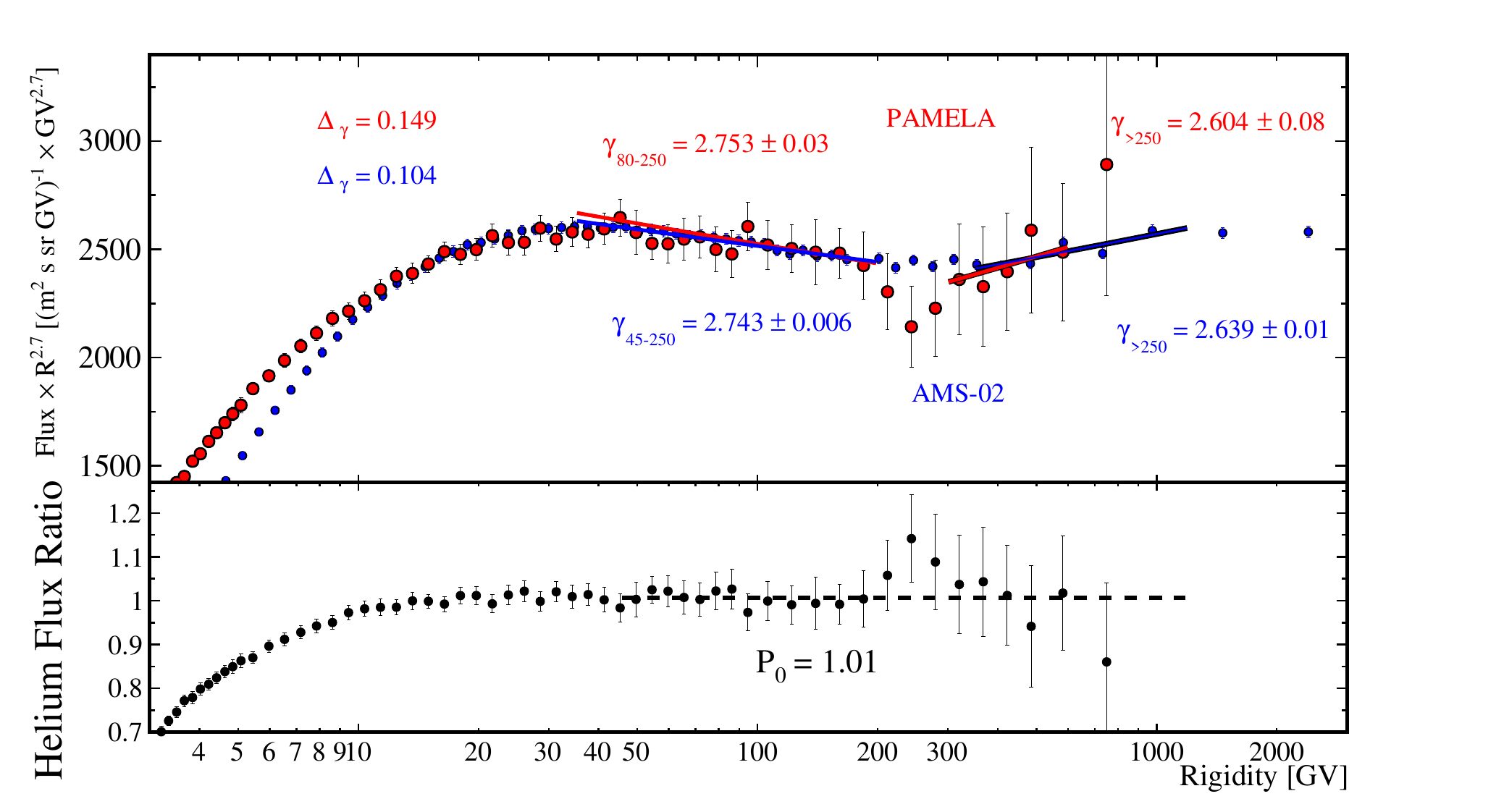}
\label{elepos_spectralindex}                            
}\vspace{-0.3cm}
\caption{(a) Comparison of proton fluxes multiplied by $E^{2.7}$ as measured by PAMELA and AMS; 
(b) Comparison of He fluxes  as a function of energy \cite{Munini}.
}
\label{figMunini}  
\vspace{-0.4cm}
\end{center}
\end{figure}

In Figs.~\ref{pfluxb} and \ref{Hefluxb} the AMS results are compared with the most recent 
measurements by magnetic spectrometers (PAMELA \cite{PAMELApHe}, BESS-Polar II \cite{Bess-pHe}) and balloon-borne 
calorimetric experiments (ATIC \cite{ATIC}, CREAM \cite{CREAMpHe}).  Systematic errors dominate in the whole rigidity range, the statistical error being less than 1\%. The total $p$ flux error is 4\% at 1 GV, $<$2\% between 2 and 350 GV, and  6\% above 1.1 TV.\\
A quantitative comparison of AMS-02 and PAMELA data was recently presented \cite{Munini}. As shown in Fig.~\ref{figMunini} an unprecedented good agreement, both in the normalization and the shape of the spectrum, is achieved in the energy range from 10 GeV to 300 GeV/n \cite{PAMELAhl}. It was pointed out that PAMELA and AMS results agree very well 
in the energy region not affected by solar modulation, i.e. at rigidities $>$30 GV, the average  
ratio between the fluxes measured by the two experiments is 0.988 for $p$ and 1.036 for He. 
It is the first time in the history of CRs that the absolute fluxes reported by different experiments differ at the percent level.
Another remark  is that, unless exploiting the transition radiation detector to extend the range of energy measurement for 
$Z\ge2$ nuclei, as proposed in \cite{Obermeier}, AMS cannot push the measurement of
$p$ and He spectra beyond its MDR limited to a few TeV/n.
Therefore, the next accurate measurements of $p$ and He fluxes over the spectral break and extending up to hundreds of TeV
are likely to come from future calorimetric experiments (see section \ref{s:futurexp}).\\

These new results suggest the need to revise the standard model of GCRs seeking an explanation for the origin of the spectral hardening.
The current paradigm is based on diffusive shock acceleration (DSA) of charged particles 
in supernova remnants (SNRs), with an injection spectrum described by a power-law in energy $E^{\alpha}$,
followed by CR propagation in the ISM, with an escape time from the Galaxy proportional to $E^{\delta}$.
It predicts that  the spectra of primary CRs observed at Earth follow
a power-law  $E^{\gamma}$  with 
spectral index $\gamma=\alpha+\delta\approx $ -2.7, i.e.  steeper than the injection spectra.
The large variety of models that have been proposed to explain the observed deviation from a single power-law behaviour can be grouped into three broad classes \cite{Serpico}. 
In the first one, the hardening is interpreted as due to the
acceleration mechanism of CRs at the source.
For instance, it could be the effect of distributed acceleration by
multiple SNRs in OB associations or Superbubbles \cite{Butt, Parizot},  
or an effect of  reacceleration of CRs by weak shocks in the Galaxy  \cite{Thoudam}, 
or alternatively it could match the spectral concavity caused by the
interactions of accelerated CRs with the accelerating shock, as foreseen 
in non-liner DSA models \cite{DSA,DSA2}.
In the second group of models, 
the spectral hardening could originate from 
subtle effects of CR propagation, like 
an inhomogeneous diffusion characterized by 
a different energy dependence in different regions of the Galaxy \cite{Tomassetti1}, 
or  a non-linear coupling of CRs with the diffusion coefficient, typical of 
models in which the transport of GCRs is regulated by self-generated waves \cite{Blasi0, Nava}.
The spectral hardening might also be due to the effect of 
local young sources \cite{Thoudam2, Liu, Zhang},
or to a mixture of fluxes accelerated by 
sources with different properties (e.g. a few old local SNRs vs. a collection of young and distant SNRs) 
contributing in different energy regions of the spectra \cite{Tomassetti2}.
Different ideas were also put forward to explain the softer $p$ spectrum with respect to 
He and heavier nuclei, including
different astrophysical sources for $p$ and He \cite{Zatsepin},
or an injection in the acceleration cycle more efficient for He than for $p$ \cite{Malkov}, 
variable He and $p$ concentrations in SNR environments \cite{Drury,Ohira},
or different spallation rates for $p$ vs heavier nuclei during propagation time \cite{Blasi}. 

\section{Light and Heavy Nuclei}
\label{s:nuclei}
In the last decade, the balloon-borne experiments CREAM \cite{CREAM-heavy} and TRACER \cite{TRACER}
measured the energy spectra of the most abundant primary heavy nuclei (C, O, Ne, Mg, Si, Fe) up to $\sim$10$^{14}$ eV particle energy.
All the spectra are well fitted by single power-laws in energy with remarkably similar spectral indices
$\gamma\approx$-2.65 in good agreement with that of He in the multi-TeV region, 
implying that all these elements are likely to share the same origin and the same
acceleration and propagation processes. By fitting simultaneously the normalized energy spectra to a broken-power law, 
CREAM reported also a hint of a spectral hardening above $\sim$200 GeV/n \cite{CREAM-harden}, 
similar to that observed for $p$ and He.\\
At the moment, the statistical and systematic uncertainties of these measurements preclude a conclusive interpretation of the data.
Further accurate observations are needed to extend with high statistics the present data to higher energies 
and measure accurately a possible curvature of the spectrum and the position of the spectral break-point for individual nuclear species.\\
\begin{figure}
\begin{center}
\subfigure[]
{
\includegraphics[height=6.6cm, width=5.4cm,angle=270]{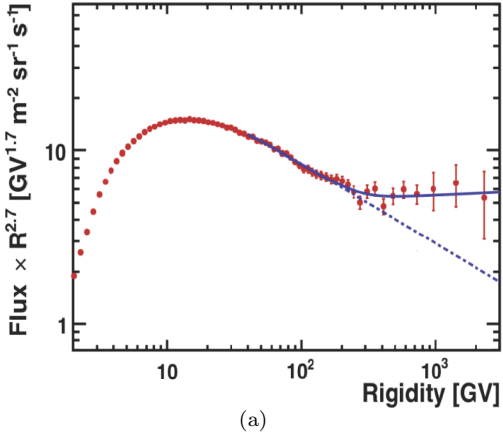}
\label{Liflux}                            
}
\hspace{10mm}
\caption{(a) The AMS Li flux multiplied by $R^{2.7}$ as a function of rigidity $R$  \cite{AMS-Li}. 
}
\label{LiCflux}
\vspace{-0.4cm} 
\end{center}
\end{figure}
So far, AMS-02 reported only preliminary measurements of the energy spectra of lithium and carbon.
Lithium, like Be and B, has a  secondary origin, being produced by spallation of heavier 
nuclei during their propagation in the ISM. Measurements of light nuclei spectra, as well as of secondary-to-primary abundance ratios,  
permit probing galactic propagation models and  constrain their parameters.
Since Li accounts for only $\sim$0.1\% of CRs, its measurement is really challenging and only scanty data from old balloon experiments were available so far. Thanks to its large geometrical factor and exposure, 
AMS-02 collected about 1.5$\times$10$^6$ Li events. The independent charge identification provided by TOF and tracker
allowed to select a pure sample of Li events, with a small contamination of heavier CR nuclei interacting in the upper part of the detector.
The differential Li flux measured between 2 GV and 3 TV \cite{AMS-Li}, 
shows an unexpected hardening above 200 GV (Fig.~\ref{Liflux}) in the same rigidity region as for $p$ and He.\\
The AMS-02 carbon spectrum \cite{AMS-C} measurement above 45 GV is still preliminary and a significant discrepancy is present between the AMS and PAMELA carbon fluxes at variance with with the general good agreement observed between their $p$ and He data. 
Preliminary results on the Li and Be fluxes from 1 to 100 GV \cite{PAMELA-LiBe}
and their isotopic composition up to 1.2 GeV/n \cite{PAMELA-LiBeIsot}
were presented at the 2015 ICRC by PAMELA as analyses in progress. \\

\begin{figure}
\begin{center}                                 
\includegraphics[height=6.0cm, width=8.4cm]{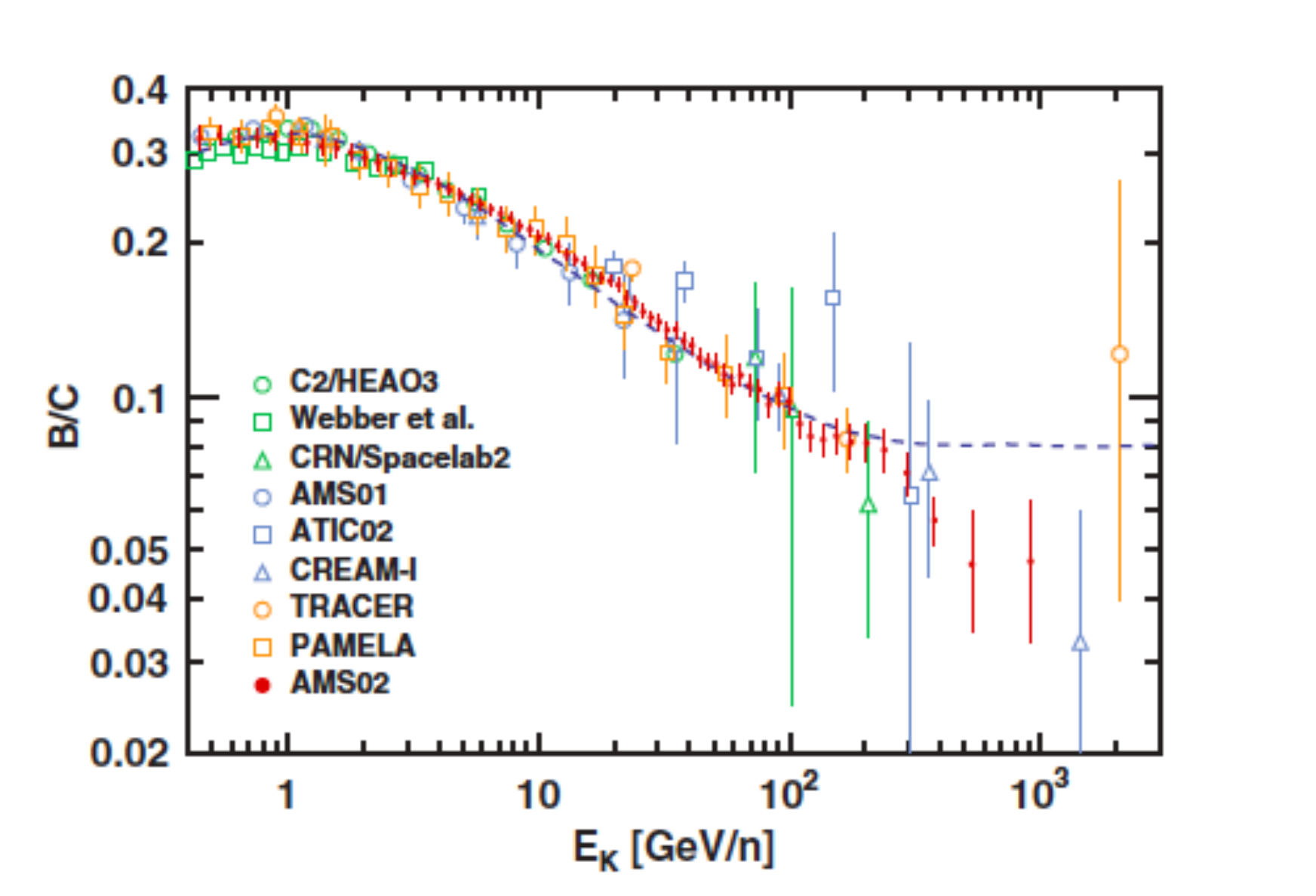}
\caption{New AMS B/C ratio compared with a compilation of earlier measurements \cite{AMS-BC}. 
}
\label{BC}   
\vspace{-0.8cm}                                                   
\end{center}
\end{figure}

\vspace{-10pt}
\subsection{Secondary-to-Primary Ratios} 

Information about the diffusion of CRs in the Galaxy can be inferred from the measurement of the secondary-to-primary ratios (e.g. B/C, sub-Fe/Fe
ratios), that allow estimation of the average amount of material traversed by CR between injection
and observation. The escape path-length $\lambda$ of charged CR from the Galaxy is known to decrease as the particle rigidity increases and to 
follow a power-law $\lambda\propto R^{\delta}$, where $\delta$ is the diffusion coefficient. 
This is a key parameter and its accurate measurement is crucial 
to derive the injection spectrum at the source by correcting the observed
spectral shape for the energy dependence of the propagation term.

Pre-AMS data on B/C ratio suggested a value of $\delta\approx$-0.6 at energies of few GeV/n, 
while at high-energy it seems to flatten to $\delta\approx$-0.4.
Measurements  by balloon experiments (CREAM \cite{CREAM-BC}, ATIC \cite{ATIC-BC}, TRACER \cite{TRACER-BC}) 
between 10$^2$ and 10$^3$ GeV/n, 
suffer from statistical limitations and significant systematic uncertainty on the correction due to the presence of
nuclei produced by the interaction of CR's with the residual atmosphere, and they do not
allow the determination of the value of $\delta$ with sufficient precision to discriminate among different propagation models. 

The new AMS data (Fig.~\ref{BC}), 
based on 2.3 million boron and 8.3 million carbon nuclei collected by AMS during the first 5 years
of operation, map the spectral shape of the B/C ratio up to about 700 GeV/n \cite{AMS-BC}.
Given this high level of precision, the theoretical uncertainties 
affecting the models become the major limitation for the interpretation of data. 

According to \cite{Genolini}, the lack of information about a  possible injection of B nuclei at the sources 
and the errors on the nuclear cross sections
seriously limit the extraction of CR diffusion parameters from B/C data.
In \cite{Kunz}, the transport parameter space is scanned 
to find solutions for the propagation equations fitting simultaneously pre-AMS B/C and other CR data.
A large number of degenerate solutions were found which 
cannot be even completely resolved by further constraints from AMS data, but 
might only be reduced with a multi-messager approach. 

Accurate measurements of B/C ratio at TV rigidities are needed to test the spectral break models. 
It has been pointed out that an error $<$10\% on the B/C ratio at 1 TeV/n would allow to discriminate between different classes of models \cite{Kunz}.
A propagation induced break would imply secondary spectra with a more pronounced break than primary ones, while no feature is expected in secondaries/primaries ratio if the break origin is in the source \cite{Serpico}.
The significant spectral break observed by AMS for Li seems to point to the first scenario, 
but more data on other nuclei are needed to clarify the picture.
However, in \cite{Aloisio}, the authors observe that  B/C ratio above 100 GeV/n is somewhat higher than the prediction 
of a non-linear CR Galactic transport model used to explain the spectral break of $p$ and He fluxes.
That could indicate a contribution of secondary production at the source that may become
significant at TeV/n scale, thus further complicating the possibility of disentangling
source and propagation effects.

\begin{figure}[!]
\begin{center}
\subfigure[]
{
\includegraphics[height=6.6cm, width=5.4cm,angle=270]{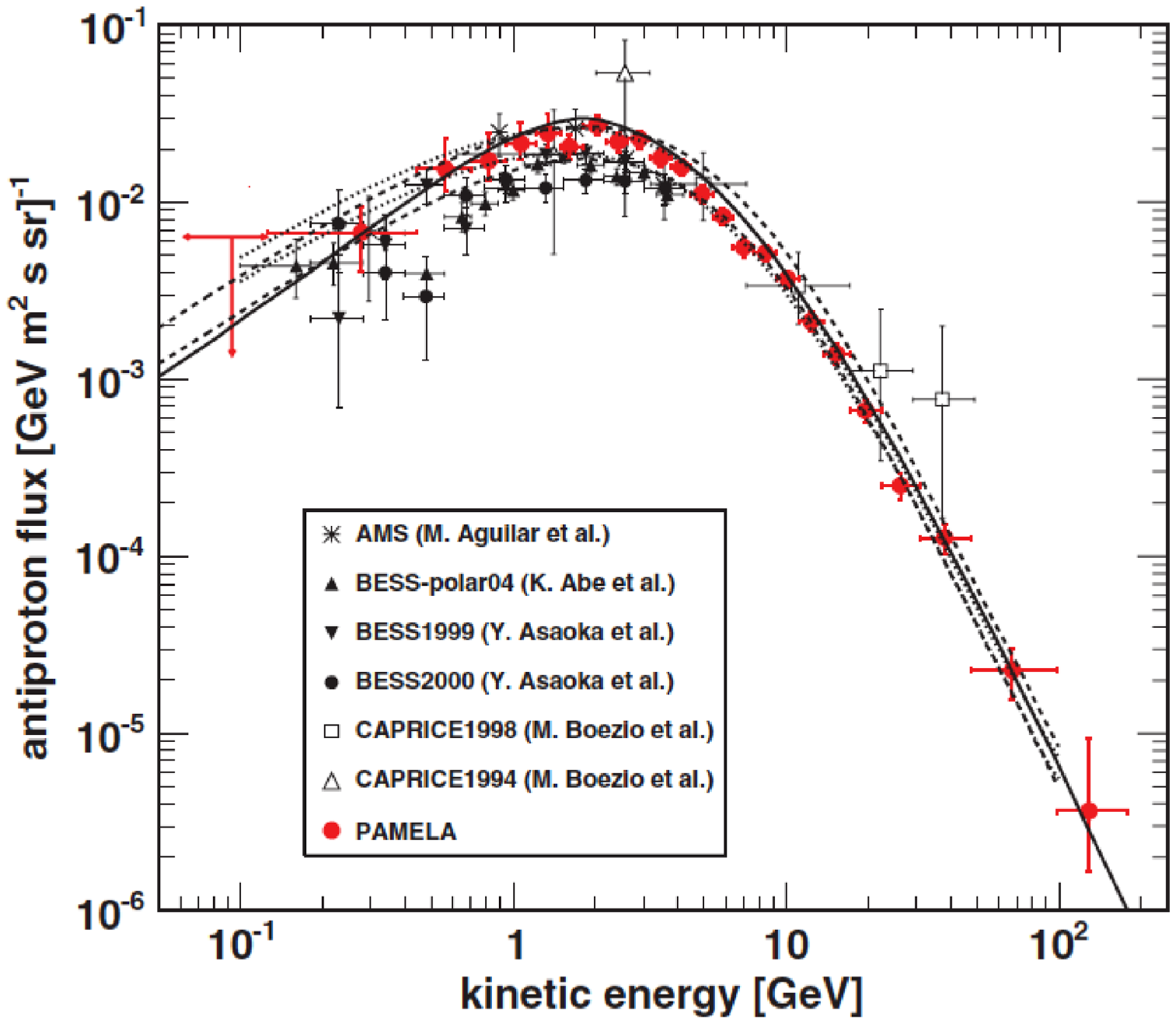}
\label{antip_a}                            
}
\hspace{8mm}
\subfigure[]
{
\includegraphics[height=6.8cm, width=5.4cm,angle=270]{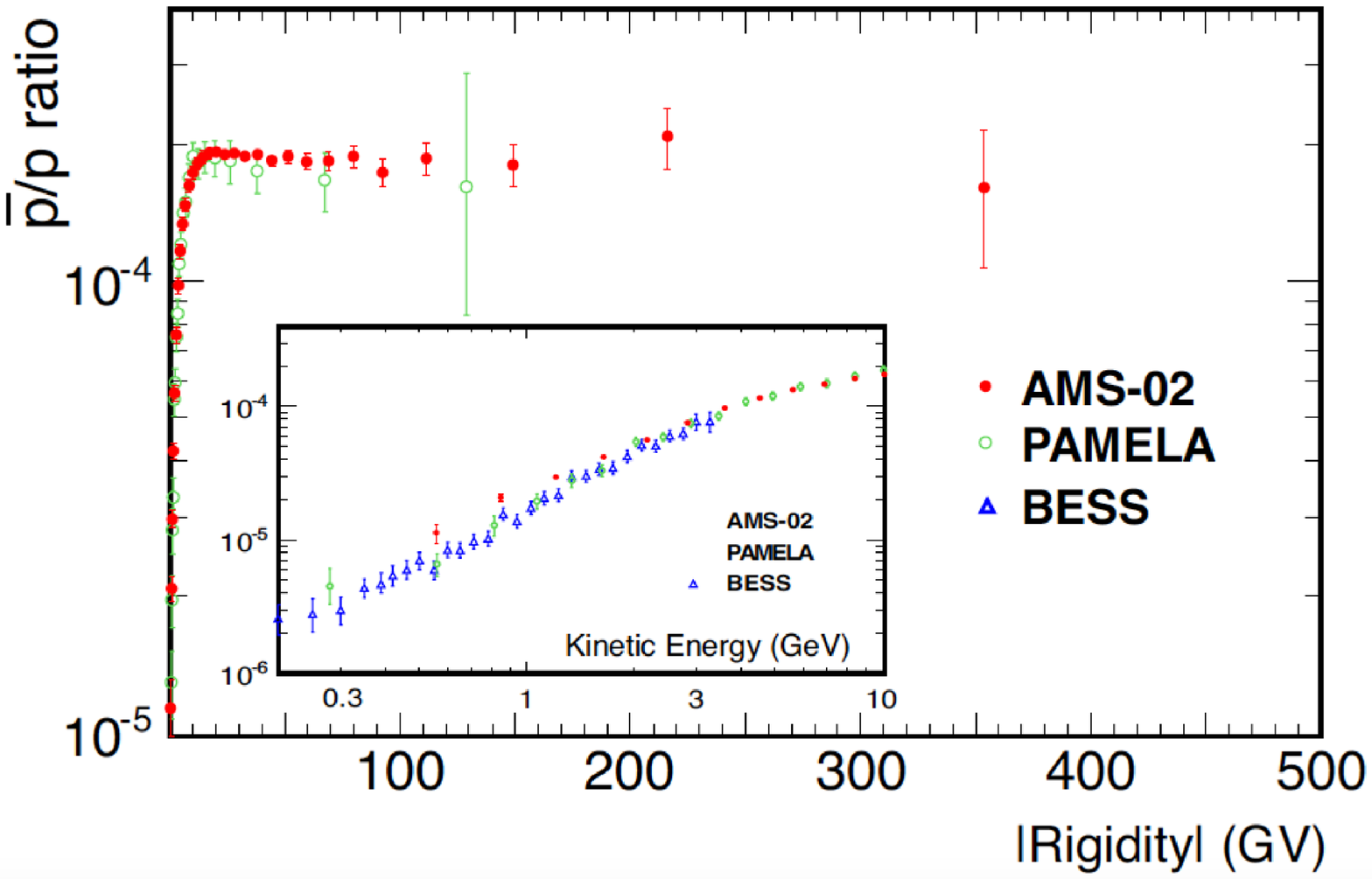}
\label{antip_b}                            
}
\caption{(a)  $\bar{p}$  flux measured by PAMELA, compared with previous results and 
theoretical calculations for a pure secondary $\bar{p}$ production \cite{PAMELA-antip}.
(b) AMS $\bar{p}/p$  ratio up to 450 GV \cite{AMS-posantipICRC}, 
compared with PAMELA \cite{PAMELA-antip} and BESS-Polar \cite{BESSPOLARII-antip} results. 
An expanded view of the low-energy region is shown in the inset plot.
}
\label{antip}
\vspace{-0.5cm}
\end{center}
\end{figure}

\vspace{-10pt}
\subsection{Trans-iron nuclei}
Elements above Fe are extremely rare in the cosmos as compared to light elements.
The abundances of elements with atomic number from Z=30 to Z=38 were measured by the Trans-Iron Galactic Element Recorder (TIGER) \cite{TIGER}
while the  ratios of $^{22}$Ne/$^{20}$Ne, $^{58}$Fe/$^{56}$Fe, and C/O  \cite{ACECRISOLD}
were provided by the Cosmic Ray Isotope Spectrometer (CRIS) on the Advanced Composition Explorer (ACE) operating at the Lagrangian point L1.
As supernovae (SN) explode preferentially in groups of massive stars, known as OB associations,
DSA might occur inside an interstellar medium of solar-system composition enriched with freshly synthesized materials from previous SN explosions
or Wolf-Rayet star ejecta. The observations of ACE support the idea of OB associations as CR acceleration sites.
Another piece of evidence supporting the OB associations model for GCR origin, 
comes from the first measurement of a primary `cosmic-ray clock', as reported by ACE-CRIS \cite{ACECRIS}.
The latter can provide an excellent elemental and isotopic separation by
measuring the dE/dx and total energy of CRs stopping in a stack of silicon detectors. 
Based on 16.8 years of collected data, CRIS was able to detect 15 well-resolved (with $\sim$0.24 mass resolution) 
events of the rare radioactive $^{60}$Fe isotope in a sample of 2.95$\times$10$^5$  $^{56}$Fe nuclei, measured
in the 240-470 MeV/n energy interval.
A detailed analysis demonstrated that at most 2 out of 15 $^{60}$Fe candidates could have been produced by interstellar fragmentation of Ni isotopes or 
could be spill-over from $^{58}$Fe. The remainig $^{60}$Fe  are likely to be primary nuclei. 
Since $^{60}$Fe undergoes $\beta^-$ decay with a half-life of 2.62 Myr, the fact that we observe this radioisotope near Earth 
implies that the time $T$ elapsed between nucleosynthesis and CR acceleration is $\le$10 Myr.
Moreover, a lower limit  of 10$^5$ yr is inferred for $T$ from the absence of $^{59}$Ni in CRs, as previosly reported by CRIS \cite{ACECRISOLD2}.
By combining these  two limits on $T$, one could conclude that CRs are not accelerated by the same SN in which they are synthesized, 
but rather CR acceleration might take place in  regions, like OB associations,  where at least two nearby supernovae occur within a few Myr.

\section{Antiprotons}
\label{s:antiproton}
Pioneer measurements of antiprotons were affected by a significant background contamination due to an inadequate particle identification
in the presence of a large background. The latter is dominated by electrons and, in the case of experiments performed with balloons at small atmospheric depths, by secondary pions and muons produced in the residual atmosphere.
In the 90's, magnetic spectrometers measured the antiproton flux below 10 GeV and only CAPRICE98 \cite{CAPRICE98_pbar} reached few tens of GeV. 
BESS95+97  showed that the $\bar{p}$ spectrum features a peak at 2 GeV while the flux is decreasing at lower energies as expected from the kinematics of $\bar{p}$ production in the ISM \cite{BESS_9597}. The observations confirmed that the secondary component is dominant in the CR $\bar{p}$ flux. However, a possible  excess at energies below the secondary peak was not ruled out by BESS95+97 data and it could originate from cosmologically primary sources of $\bar{p}$ as, for instance, the evaporation of primordial black holes (PBH). This exotic component was later excluded by BESS-Polar II precise measurement between 0.17 and 3.56 GeV near solar minimum in December 2007 \cite{BESSPOLARII-antip}. 


The first observations of the $\bar{p}$ flux (Fig.~\ref{antip_a}) and 
$\bar{p}/p$ flux ratio extending to high energy (180 GeV) were performed by PAMELA \cite{PAMELA-antip}.
AMS-02 presented a preliminary measurement of the $\bar{p}/p$ ratio in 2015, based on the identification of 2.9$\times$10$^5$ $\bar{p}$ selected in the rigidity range 1-450 GV \cite{AMS-posantipICRC}. The data showed that the $\bar{p}/p$ ratio is almost constant above a rigidity of 100 GV (Fig.~\ref{antip_b}).\\
This unexpected behaviour at high energy triggered several speculations about a possible primary antiproton component originating from the annihilation of DM particles.

\begin{figure}[b]
\begin{center}
\subfigure[]
{
\includegraphics[height=6.6cm, width=5.4cm,angle=270]{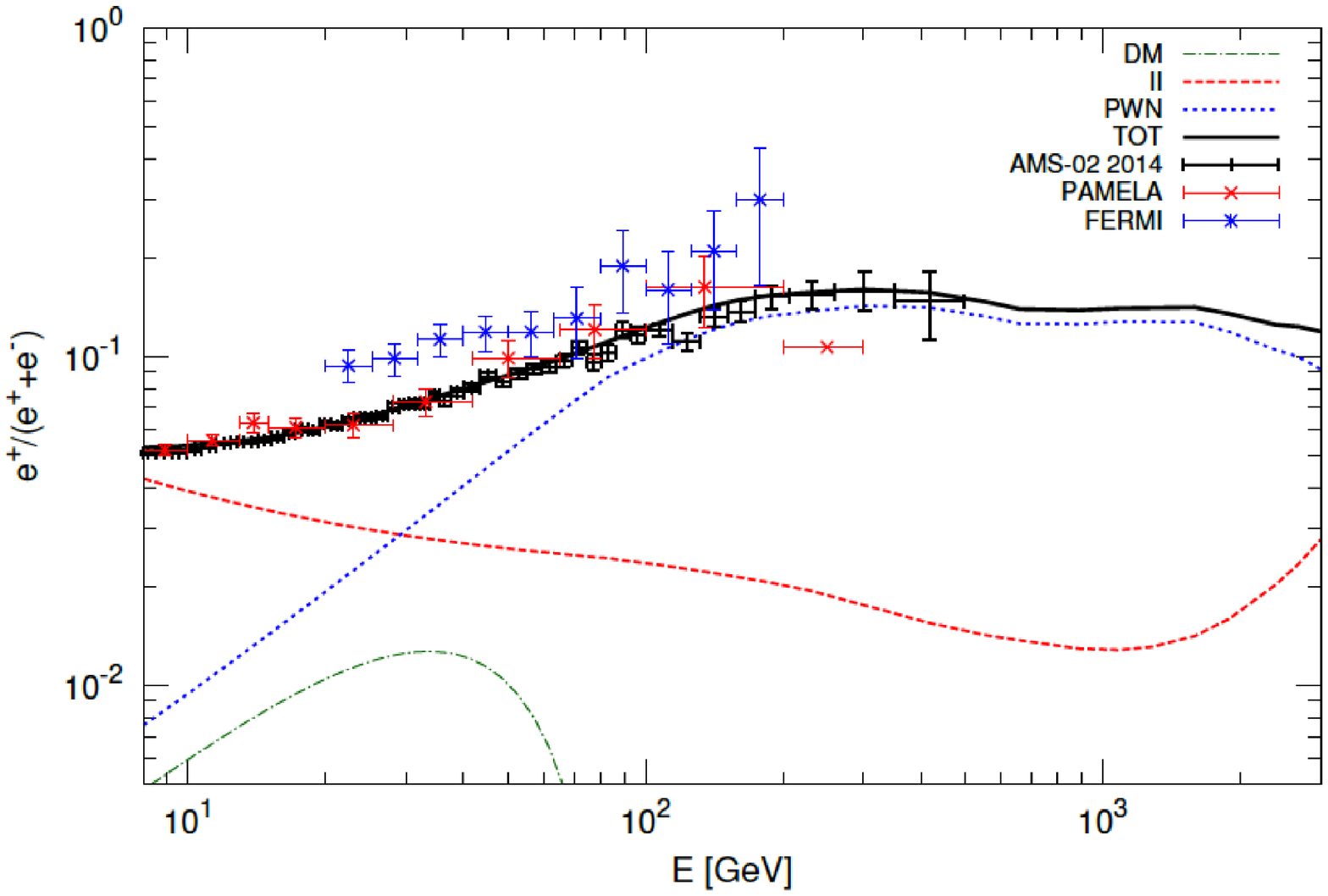}
\label{posfrac_dimauro}                            
}
\hspace{7mm}
\subfigure[]
{
\includegraphics[height=6.6cm, width=5.4cm,angle=270]{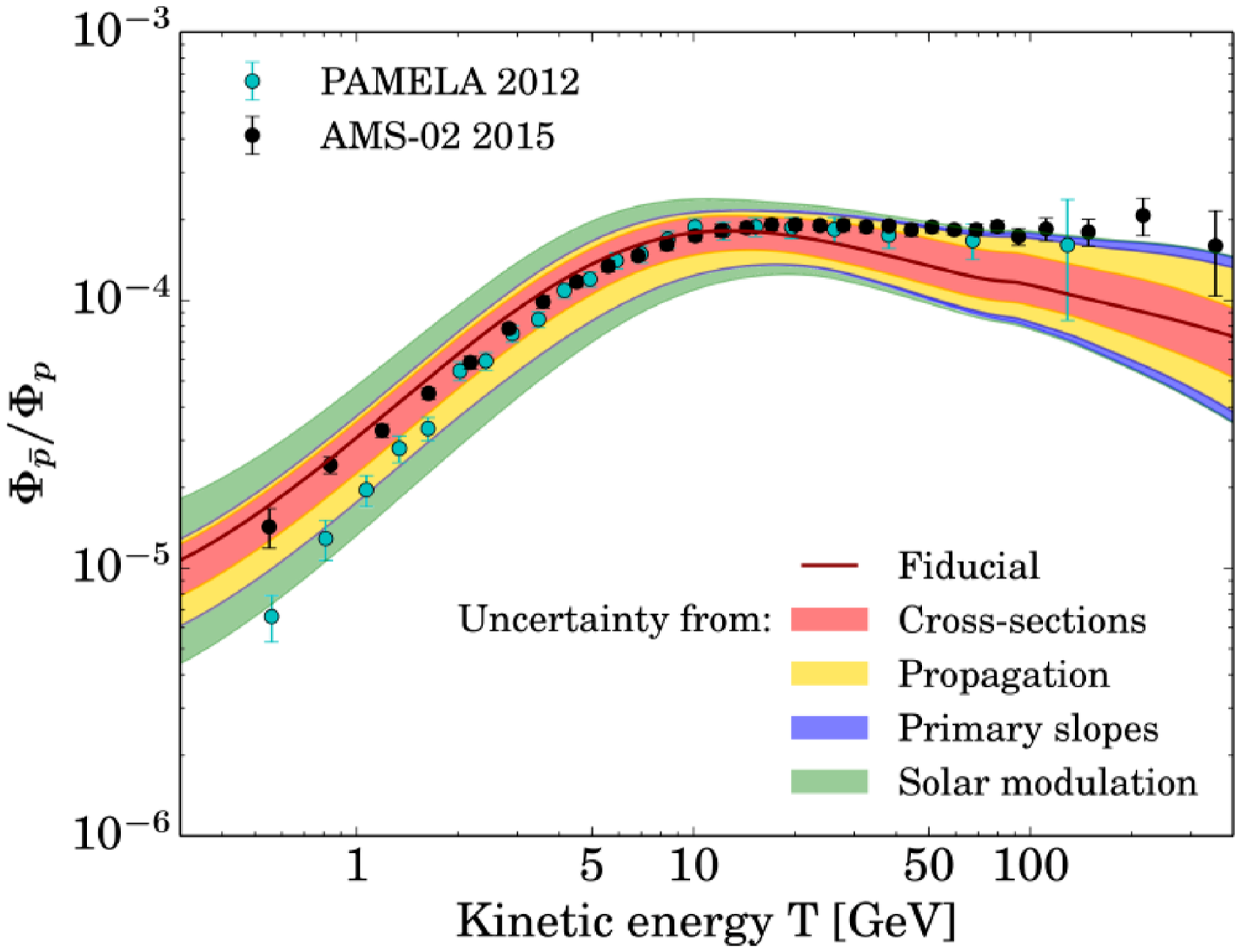}
\label{antip_boudad}                            
}
\caption{(a) Prediction for the positron fraction calculated in \cite{ICRCDiMauro}, 
including contributions from  secondaries, PWN and DM annihilating into the $\mu^+\mu^-$ channel. (b) The combined total uncertainty on the secondary $\bar{p}/p$ ratio calculated in \cite{ICRCGiesen}, superimposed to PAMELA and AMS data.
}
\label{theory}
\end{center}
\end{figure}
After a misleading comparison of AMS data with old secondary production predictions (e.g. \cite{Donato}), 
calculations of the secondary $\bar{p}/p$ flux ratio have been reassessed \cite{Donato2,ICRCGiesen,Cowsik1,Kappl,Evoli}
using updated propagation models and by taking into account more recent data on $\bar{p}$ production cross sections,
diffusive reacceleration, solar modulation, the hardening of $p$ and He spectra at high energy, 
an updated value of the diffusion coefficient consistent with recent B/C data, and additional effects
like $\bar{p}$ energy losses that had been neglected so far. The majority of the models suggest that the flat $\bar{p}/p$ ratio at high energy is consistent with pure secondary production within the error band of the updated predictions (for a review see for instance \cite{Cowsik2}). 
However, given the current uncertainties, mainly on propagation parameters, 
a possible $\bar{p}$ contribution from DM cannot be ruled out.


\section{New Cosmic-Ray Explorers}
\label{s:futurexp}
At the time of writing, a new generation of instruments has been successfully placed in Low Earth orbit (NUCLEON, CALET, DAMPE), while new experiments are planned to start data taking soon (ISS-CREAM, CSES). Other missions are at a proposal stage (HNX, HELIX, GAPS) or under conceptual design and development (GAMMA-400, HERD).
The only two magnetic spectrometers in orbit today are PAMELA and AMS-02 and it is likely that considerable time will elapse before a new space instrument capable of providing sign-of-charge discrimination will be built. In any case, an MDR limited to a few TV implies a non competitive momentum resolution for electrons with respect to the superior energy resolution of calorimeters in this range. Therefore, instruments like CALET, NUCLEON, DAMPE, ISS-CREAM have been designed as all-calorimetric with sufficient depth to provide shower containement for TeV electrons and a hadronic energy deposit that can be related to the energy of the primary particle up to several hundred TeV and, given a sufficient exposure, can reach the PeV scale. The geometric factors are of the order of a thousand cm$^{2}$sr (to be compared with 20 cm$^{2}$sr of PAMELA) and mission lifetimes of several years.  They all implement charge identification techniques to study individual spectra of CR nuclei up to their maximum energy reach and electrons up to several TeV. \\
Other missions like CSES, HNX, HELIX, GAPS are more focused on specific observation targets and will be briefly mentioned in the following. \\


NUCLEON is a Russian satellite-borne instrument designed to measure the energy spectra of individual CR nuclei from $Z$=1 to $Z$=30 in the energy range from 100 GeV up to 1 PeV using a traditional ionization calorimeter complemented by an independent technique, known as Kinematic Lightweight Energy Method (KLEM), whereby the energy of the primary particle is inferred from the pseudo-rapidities of secondaries produced in a carbon target (and tungsten absorbers) and measured by a Si-microstrip tracker that records their spatial density. The total thickness is $\sim$16 $X_0$. The sampling calorimeter can also measure electrons in the energy range 100GeV -- 3TeV.  NUCLEON was successfully launched on a RESURS-P satellite on December 26, 2014. The effective geometric factor is more than 0.2 m$^2$sr for nuclei and 0.06 m$^2$sr for electrons \cite{NUCLEON}. The mission is planned to take data for at least 5 years.\\

CALET is a space mission led by the Japanese Space Agency (JAXA) with the participation of the Italian Space Agency (ASI) and NASA. It was launched on August 19, 2015 by the Japanese H-II and delivered to the ISS by the HTV-5 Transfer Vehicle, where it was installed on the Japanese Experiment Module Exposure Facility (JEM-EF).
CALET is an all-calorimetric instrument \cite{Torii2011, Torii2013, Mar2013} designed to achieve a large proton rejection capability ($>$10$^{5}$) with a fine grained imaging calorimeter (IMC) followed by a total absorption calorimeter (TASC). The overall thickness of CALET at normal incidence is 30 X$_{0}$ and $\sim$1.3 proton interaction length ($\lambda_{I}$). 
The charge identification of individual nuclear species is performed by a two-layered hodoscope of plastic scintillators (CHD) at the top of the apparatus providing a measurement of the charge Z of the incident particle over a wide dynamic range ($Z=1$ to $\sim40$) with sufficient charge resolution to resolve individual elements \cite{Shimizu2011, Mar2011} and complemented by a redundant charge determination via multiple dE/dx measurements in the IMC. 
The latter is a sampling calorimeter longitudinally segmented into 16 layers of scintillating fibers readout individually.
It can image the early shower profile in the first 3 X$_{0}$ and reconstruct the incident direction of cosmic rays with good angular resolution (0.1$^\circ$ for electrons and better than 0.5$^\circ$ for hadrons).
%
The TASC is a 27 X$_0$ thick homogeneous calorimeter with 12 alternate X-Y layers of lead-tungstate (PWO) logs.\\
%
  CALET main science objectives include the exploration of the electron (+positron) spectrum above 1 TeV whose shape might reveal the possible presence of nearby sources of acceleration. With excellent energy resolution (better than 2\% for electrons above 100 GeV,
proton rejection capability $>$ 10$^{5}$) and low background contamination, CALET will search for possible signatures of dark matter in the spectra of both electrons and gamma rays. High precision measurements of the energy spectra, relative abundances and secondary-to-primary ratios of cosmic nuclei from proton to iron will be carried out as well as the detection of trans-iron elements. Deviations from a simple power-law, as reported by CREAM, PAMELA and AMS-02 in proton and He spectra, will be studied with high accuracy in the region of a few hundred GeV and extended to the multi-TeV region and to heavier nuclei. Gamma-ray transients are detected by a dedicated Gamma-ray Burst Monitor (GBM). The geometrical factor is 0.12 m$^2$sr and the total weight is 613 kg. \\

The ISS-CREAM payload inherits from the CREAM balloon instrument that successfully accumulated data for 161 days in six flights over Antarctica and will be installed on the ISS JEM-EF for a 3 year mission \cite{ISSCREAM}.
The increase of the exposure by nearly an order of magnitude will allow to extend the measurements of the energy spectra of CR nuclei to hundreds of TeV and $p$ and He up to few PeV.
The ISS-CREAM payload is scheduled to be delivered to the ISS by SpaceX in mid 2017.
The instrument  ($\sim$1300 kg) is optimized for hadronic measurements with a 0.5 $\lambda_I$ carbon target to induce the inelastic interaction of the incoming nuclei followed by a sampling calorimeter (20 X$_{0}$)  with 20 layers of alternating tungsten plates and scintillating fibers, 
providing energy measurement, particle tracking and trigger and four layers of Si pixels for particle charge identification. Electron measurements will also be possible with the help of top and bottom plastic scintillator counters and a boronated scintillator detector for $e/p$ separation.\\

DAMPE  is a satellite experiment promoted by the Chinese Academy of Sciences and built in collaboration with institutions from Italy and Switzerland \cite{DAMPE}. Launched on December 2015, DAMPE is designed to measure electrons and
photons from 5 GeV to 10 TeV with excellent resolution (1.5\% at 100 GeV) thanks to its BGO imaging calorimeter. It will
search for possible DM signatures and study the spectra of CR nuclei in the range 10 GeV to 100 TeV with individual element separation.
The total thickness of the calorimeter is 31 $X_0$ and 1.6 $\lambda_I$. The instrument is equipped at the top with plastic scintillator strips, as charge detector for CRs and anti-coincidence for photons, and by a silicon-tungsten tracker (STK) providing track reconstruction, photon detection and redundant CR charge measurement. A neutron detector with boron-doped plastic plates follows the calorimeter to improve the $e/p$ discrimination.
The geometrical acceptance is $\sim$0.3 m$^2$sr for $e^-$ and $\gamma$, and $\sim$0.2 m$^2$sr for hadrons.\\

CSES is a chinese-italian space mission
designed to study electromagnetic anomalies and ionosphere perturbations as well as the precipitation of low energy  electrons
trapped within the Van Allen Belts and their possible correlation with earthquakes.
The HEPD (High Energy Particle Detector) will also study the low-energy component of CRs of solar and galactic origin.
It consists of two layers of plastic scintillators for trigger,  
two planes of double-side Si microstrip detectors to measure the incident particle direction, 
a calorimeter of plastic and LYSO crystal scintillators, 
and lateral  veto system for a total weight of 35 kg.
It can measure electrons from 1 to 200 MeV and protons from 30 to 200 MeV with very good separation,
energy resolution $<10$\% at 5 MeV and an angular resolution of $\sim$8$^\circ$ \cite{CSES}.\\

GAMMA-400  is a proposed next generation gamma-ray space observatory \cite{GAMMA400}.
Though specifically designed to
measure gamma-ray sources with unprecedented accuracy, the  instrument also 
includes an innovative homogeneous and isotropic calorimeter made of cubic crystals, 
featuring a depth of 54 X$_0$ or
2.5 $\lambda_I$ when detecting laterally incident particles. 
The resulting improved  shower containment and the large acceptance (a few m$^2$sr) would allow to measure
 CR nuclei up to the ``knee'' with very good energy resolution. \\

HERD  is a next generation gamma-ray space observatory to be deployed onboard 
China’s Space Station and planned for 2023-2025 \cite{HERD}. The core of HERD is a three dimensional LYSO crystal homogeneous calorimeter with fiber readout sensitive to incoming CR radiation on five surfaces. Because of this capability, the effective geometric factor of HERD is increased by an order of magnitude compared to previous instruments with similar weight, size and power consumption. According to a preliminary design, the micro silicon strip tracker includes a top tracker equipped with silicon strips and thin tungsten foils (the latter acting as converter for gamma-rays) and four sides instrumented with 3 layers of silicon strip detector with dimensions 100 cm $\times$ 70 cm each. The top tracker provides charge identification, trajectory measurement, back scatter rejection and imaging of the early shower development for gamma  and  electron.  
In addition to its sensitivity to high energy electrons and gamma rays, HERD will study primary cosmic-ray composition and spectrum to the PeV range. \\

HNX  (Heavy Nuclei eXplorer)) is a NASA proposed space mission inheriting from the experience of the balloon-borne SuperTIGER \cite{HNX}. 
It includes complementary active detectors (silicon strips and Cherenkov counters with acrylic and silica-aerogel radiators), 
and passive glass tiles to achieve 50 m$^2$sr geometric factor, required to measure nuclei up to the Actinides (Th, U, Pu) 
which can provide information about
the absolute age of ultra-heavy GCRs since nucleosynthesis.
HNX is designed to be accomodated in the DragonLab Capsule for a 2 year mission. \\

HELIX (High Energy Light Isotope eXperiment) is a balloon-borne 
superconducting magnet spectrometer, complemented with
a Ring Imaging Cerenkov Detector and a Time-of-Flight system,
to measure with good statistics the $^{10}$Be/$^{9}$Be abundance ratio 
with  0.25 amu mass resolution \cite{HELIX} up to 10 GeV/n. 
This measurement can provide very important information 
on the mean containment lifetime of CRs in the Galaxy. \\

GAPS (General Antiparticle Spectrometer) is a large-acceptance balloon-borne experiment 
that will measure low-energy ($<$1 GeV/n) cosmic antideuterons and $\bar{p}$ for indirect DM searches. 
Antimatter particles are identified by the characteristic X-rays emitted by exotic atoms formed when they are stopped in 
a stack of Si(Li)  detectors and by the signature of the production of a $\pi$ or $p$ in a star pattern
when subsequently they  annihilate. This new detection technique is almost background free \cite{GAPS}. \\

\vspace{-0.3cm}
\section{Conclusions}
A new era of precision measurements of CRs has just started. Instruments with a much larger collecting power than their predecessors are now in orbit or expected to become operational in the near future.  They are able to provide high statistics measurements of cosmic rays with excellent instrumental performances and unprecedented capabilities to control the systematic errors. New unexpected findings (including the momentum dependence of spectral indices, the striking similarity of the positron spectrum with proton's, the energy dependence of the positron fraction) are waiting for a coherent theoretical 
framework describing the mechanism(s) of acceleration of galactic cosmic rays, the nature and composition of their
sources, their propagation in the interstellar madium.\\
In our opinion, the hope of a breakthrough in our understanding of this field may not be unresonable in the relatively near future.


\end{document}